\documentclass{mn2e}
\usepackage{epsfig}

\usepackage{graphicx}

\newif\ifAMStwofonts
\AMStwofontstrue

\title[The K-band Hubble diagram of sub-mm galaxies and 
hyperluminous galaxies]
{The K-band Hubble diagram of sub-mm galaxies and 
hyperluminous galaxies}
\author[Serjeant, et al.]
  {Stephen Serjeant$^1$, Duncan Farrah$^{2}$, 
James Geach$^{3}$, Toshinobu Takagi$^1$, 
\vspace*{0.3cm}\\
{\LARGE\rm 
Aprajita Verma$^4$, Ali Kaviani$^{3}$, 
Matt Fox$^{3}$}\\
\vspace*{0.1cm}\\
$^1$Centre for Astrophysics \& Planetary Science, School of Physical
Sciences, University of Kent, Canterbury, Kent CT2 7NR, UK\\
$^2$SIRTF Science Center, Jet Propulsion Laboratory, California
Institute of Technology, Pasadena, 91125, USA\\
$^3$Astrophysics Group, Imperial College London, Blackett Laboratory,
Prince Consort Road, London SW7 2BW\\
$^4$MPI fuer extraterrestrische Physik, Giessenbachstr. 1, D-85748,
Garching, Germany
}
\date{Received 2003}
\pubyear{2003}

\begin{document}


\input BoxedEPS.tex
\SetEPSFDirectory{./}

\SetRokickiEPSFSpecial
\HideDisplacementBoxes

\label{firstpage}

\maketitle

\begin{abstract}
We present the $K$-band Hubble diagrams ($K-z$ relations) of
sub-mm-selected galaxies and hyperluminous galaxies (HLIRGs). 
We report the discovery of a remarkably tight $K-z$ relation
of HLIRGs, 
indistinguishable from that of the most luminous radiogalaxies.
Like radiogalaxies, the HLIRG K-z relation at $z\stackrel{<}{_\sim}3$
is consistent with a passively evolving 
$\sim3L_*$ instantaneous starburst starting from a redshift of $z\sim10$. 
In contrast, many sub-mm selected galaxies are $\stackrel{>}{_\sim}2$
magnitudes fainter, and the population has a much larger
dispersion. We argue that dust obscuration and/or a larger mass range 
may be responsible for this scatter. 
The 
galaxies so far proved to be hyperluminous may have been biased
towards higher AGN bolometric contributions than sub-mm-selected galaxies 
due to the $60\mu$m
selection of some, so the location on the $K-z$ relation
may be related to the presence of the most massive AGN. Alternatively,
a particular host galaxy 
mass range may be responsible for both extreme star formation and the
most massive active nuclei. 
\end{abstract}


\section{Introduction}\label{sec:introduction}
Hyperluminous galaxies (HLIRGs, $L>10^{13}L_\odot$, as distinct from
the less-luminous ultraluminous population with $L=10^{12-13}L_\odot$), 
were first identified from follow-ups of the IRAS mission
(e.g. Kleinmann et al. 1988, Rowan-Robinson et al. 1991). 
Gravitational lensing was found to be responsible for some of 
the extreme
luminosity of at least one HLIRG, IRAS F10214+4724 
(Graham \& Liu 1995, Serjeant et
al. 1995, Broadhurst \& Lehar 1995, Eisenhardt et al. 1996), but 
subsequent HST imaging of more HLIRGs showed no further lens
candidates (Farrah et al. 2002a). The morphologies were 
found to be diverse, from interacting to quiescent. 
Although active nuclei have been found in all HLIRGs to date, the
enormous gas and dust masses (e.g. Downes et al. 1993, Clements et
al. 1992, Farrah et al. 2002b) are indicative of violent, possibly
bolometrically-dominant, star formation. 
By fitting multi-wavelength photometry of HLIRGs, several authors
have found comparable bolometric contributions from star formation and
active nuclei in many HLIRGs 
Hyperluminous galaxies appear to be a population of 
galaxies undergoing their major star formation episode (Rowan-Robinson
2000), but at an epoch in which AGN activity is also present
(e.g. Rowan-Robinson 2000, Farrah et al. 2002b, Verma et  
al. 2002). 
The sub-mm detections of radiogalaxies (Archibald et al. 2001)
and quasars (e.g. Priddey et al. 2003) further
supported a link between violent star formation and AGN activity,
though quasar-heated dust has also been raised as a possibility
(Willott et al. 2002). In this paper we will present further evidence
for a link between AGN activity and extreme star formation, using the
$K$-band Hubble diagram. 

The tight dispersion in the 
$K$-band Hubble diagram ($K-z$ relation) of radiogalaxies
has long been held to suggest a high
formation epoch for radiogalaxy hosts (Lilly \& Longair
1984). Redshifted emission line contributions (Eales \& Rawlings 1993)
complicate the interpretation at redshifts $z>2$, but largely 
only for the most
luminous radiogalaxies (e.g. Jarvis et al. 2001). The current
consensus is that the tight $\pm0.5$ magnitude 
dispersion in the radiogalaxy $K-z$
relation persists at $z>2$, and is still consistent with a passively
evolving stellar population with a formation
epochs at $z>2.5$. There is also a weak correlation of $K$-band
luminosity with radio luminosity at any epoch (e.g. Willott et
al. 2003) which has been attributed to mutual correlations with
central nuclear black hole masses. 
Furthermore, the host galaxies of radio-loud AGN tend to be restricted
to a more luminous population than their radio-quiet counterparts
(Dunlop et al. 2003a), suggesting that it is only the most massive
($>10^9M_\odot$) nuclear black holes which give rise to radio-loud
AGN. Finally, the similarity of the $K$-band morphologies of
sub-mm-selected galaxies to those of high-$z$ radiogalaxies, the high
star formation rates in sub-mm galaxies (sufficient to assemble a
giant elliptical in $\sim10^8$ years), and the presence of radio-loud
AGN in local ellipticals has suggested to some authors that both
high-$z$ radiogalaxies and sub-mm selected galaxies are the
progenitors of the most massive spheroids (e.g. Dunlop 2002, Scott et
al. 2002). 



In this paper we report the discovery of a remarkably tight $K-z$
relation of HLIRGs, and the surprising lack of a tight $K-z$ relation
for coeval sub-mm-selected galaxies.  
Section \ref{sec:method} describes the compilation
of $K$-band magnitudes, and the $K-z$ relations are presented in
section \ref{sec:results}. Section \ref{sec:discussion} places the
results in the context of other high-$z$ populations, and discusses
the physical implications and possible applications of this relation. 
Throughout this paper, ``quasars'' are taken to mean objects with
broad ($\stackrel{>}{_\sim}2000$ km s$^{-1}$) unpolarised emission
lines, regardless of the presence or absence of a host galaxy in
imaging data, and we assume $\Omega_{\rm M}=0.3$,
$\Omega_\Lambda=0.7$ and 
$H_0=70$ km s$^{-1}$ Mpc$^{-1} = 100h$ km s$^{-1}$ Mpc$^{-1}$.  
In this cosmology, a minority of the 
hyperluminous galaxy compilation of
Rowan-Robinson (2000) slip just below the hyperluminous threshold, and
others attain hyperluminous status, 
but for consistency with previous works we restrict ourselves to this
compilation. This choice does not affect the results in this paper. 

\section{Method}\label{sec:method}
 
\begin{table}
\begin{tabular}{llllll}
Name             & $z$ & $\log_{10}L_{\rm bol}$ & $K$ & Aper & Notes\\
IRAS F00235+1024 & 0.58 & 13.15 & 17.19 & $3''$ & (4)\\  
SMM J02399-0136   & 2.803 & 13.08 & 18.79 & $3''$ & (1,7)\\ %
4C41.17         & 3.8  & 13.12 & 19.6  & $4''$ & (11)\\ %
IRAS 09104+4109  & 0.44 & 13.24 & 15.41 & $3''$ & (4)\\ %
IRAS F10214+4724 & 2.286 & 13.54 & 20.10 & $\sim3''$ & (2,9)\\ 
IRAS F12514+1027 & 0.30 & 13.00 & 13.48 & $10''$ & (5)\\ %
SMM J14011+0252 & 2.55 & 13.18 &  18.71 & $3''$ & (3,7)\\
IRAS F15307+3252 & 0.93 & 13.50 & 16.59 & $2''$ & (10)\\
FFJ1614+3234    & 0.710 & 13.07 & 16.6  & $3''$ &  (8)\\ 
\end{tabular}
\caption{\label{tab:kmags}
K-band magnitudes for non-quasar hyperluminous galaxies of
Rowan-Robinson (2000). Following Dunlop et al. (2003) we neglect the
corrections between $K'$, $K$ and $K_{\rm s}$ as these are smaller
than the typical photometric errors ($\sim0.1-0.2$ magnitudes). 
Bolometric luminosities are taken from Farrah et al. (2003), Farrah et
al. (2002b) and Rowan-Robinson (2000). 
Notes: (1) corrected for $\times1.3$ lens
magnification factor; (2) corrected for emission line contribution and
$\times10$ magnification factor; (3) corrected for $\times2.8$ lens
magnification; (4) photometry from the data of 
Farrah et al. 2003 (note that those authors used slightly different
photometric apertures); (5) photometry
from 2MASS (Cutri et al. 2003); (6) photometry from Eisenhardt \&
Dickinson 1992; 
(7) photometry from  
Smail, Ivison, Blain \& Kneib 2002b; 
(8) photometry from Stanford et al. 2000; 
(9) photometry
from Graham \& Liu 1995; 
(10) photometry from Liu et al. 1996; 
(11) photometry from Graham et al. 1994
}
\end{table}

%
%

\begin{table}
\begin{tabular}{lllll}
Name             & $z$ & $K$ & Aper & Notes\\
SMMJ02399-0134 & 1.06  & 17.29 & $3''$ & (7)\\
SMMJ02399-0136 & 2.803 & 18.79 & $3''$ & (7)\\
CUDSS10A       & 0.550 & 17.11 & $3''$ & (16)\\
LH850.11       & 2.610 & $>$20.6 & $4''$ & (13,21)\\
LH850.12       & 2.698 & $>$20.6 & $4''$ & (13,21)\\
LH850.15       & 2.429 & $>$20.4 & $4''$ & (13,21)\\
LH850.18       & 3.699 & $>$20.4 & $4''$ & (13,20)\\
HDF850.4       & 0.475 & 18.14 & $6''$ & (14,20)\\
HDF850.6       & 0.884 & 19.85 & $6''$ & (14,20)\\
HDF850.8       & 1.355 & 19.69 & total & (15,20)\\
SMMJ14011+0252 & 2.55  & 18.71 & $3''$ & (7)\\
CUDSS14F       & 0.660 & 16.81 & $3''$ & (16)\\
N2850.1        & 0.840 & 19.48 & $4''$ & (13,20)\\
N2850.4        & 2.376 & 18.43 & $4''$ & (13,20)\\
N2850.8        & 1.189 & 18.15 & $4''$ & (13,20)\\
SMMJ17142+5016 & 2.39  & 21.80 & total & (12)\\
\hline
SMMJ00266+1708 & 2.0-5.0 & 23.45 & $3''$ & (7,19)\\
SMMJ09429+4658 & 2.1-4.5 & 20.15 & $3''$ & (7,19)\\
LH850.1        & 2.0-3.0 & 19.8 & $4''$ & (13,19)\\
LH850.3        & 1.2-3.5 & 18.86 & $4''$ & (13,19)\\
LH850.8        & 2.4-5.2 & 18.82 & $4''$ & (13,19)\\
HDF850.1       & 3.5-4.7 & 23.5 & $1''$ & (17,19)\\
SMMJ14099+0252 & 2.6-5.1 & 21.44 & $3''$ & (7,19)\\
CUDSS14.1      & 2.0-4.6 & 19.55 & $4''$ & (18,19)\\
\end{tabular}
\caption{\label{tab:kmags_scuba}
K-band magnitudes for sub-mm selected galaxies with spectroscopic
redshifts or $\geq3$ band photometric redshifts. Notes as table
\ref{tab:kmags}, and in addition: (12) from Smail et al. 2003 and
Keel et al. 2002, and corrected for emission line contributions; (13)
from Ivison et al. 2002; (14) Barger et 
al. 2000; (15) Fernandez-Soto et al. 1999; (16) Lilly et al. 1999; 
(17) from Dunlop et al. 2003, photometric aperture is small but source
is not significantly more extended, and no account taken of lensing;
(18) Gear et al. 2000; (19) Aretxaga et al. 2003; (20) Serjeant et
al. 2003; (21) Chapman et al. 2003. Non-sub-mm selected
galaxies (e.g. HR10) that are nevertheless detected in sub-mm
photometry are excluded from this table. 
}
\end{table}

Table \ref{tab:kmags} lists the $K$ magnitudes for all non-quasar
hyperluminous galaxies in the compilation of Rowan-Robinson (2000), 
except IRAS F14481+4454 and IRAS F14537+1950 for which no data is
available. Several quasar HLIRGs have optical host galaxy
measurements (e.g. Farrah et al. 2002a), but we exclude these on the
grounds that the $K$-band nuclear contribution will
differ. 

There are great difficulties with associating $14''$ resolution
sub-mm blank-field survey galaxies with
sources at other wavebands (e.g. Serjeant et al. 2003), especially
given the non-Poissonian distribution of these and other populations
(e.g. Almaini et al. 2003). 
Most claimed spectroscopic redshifts 
report spectroscopic evidence of star formation (e.g. Chapman et
al. 2003), but one should note that 
there are precedents for
identifications of sub-mm sources changing with the advent of new data
(e.g. Dunlop et al. 2003b, Serjeant et al. 2003, Smail et al. 1999). 
The $K$ magnitudes of all published sub-mm galaxies with spectroscopic
redshifts are listed in table \ref{tab:kmags_scuba}. We also include
galaxies with $\geq3$ band photometric redshifts, though the accuracy
of these redshifts will depend on how closely the high-$z$ population
resembles local templates (yet to be determined). 

In addition to HLIRGs and sub-mm galaxies, 
we have compiled comparative data from the
literature on other, ostensibly related populations. 
We use the photometry and redshifts of non-quasar Chandra sources
from the Hubble Deep Field North Chandra $1$ megasecond catalogue
of Barger et al. (2002), which is an improvement on the inhomogeneous
compilations available to Willott et al. (2001). We also use the
$K$-band photometry and spectral classifications 
of ultraluminous infrared galaxies of Kim et al. (2002) and Veilleux
et al. (2002). 
Finally, we use the radiogalaxy photometry compilation in Willott et
al. (2003). 
Because the 3CRR flux limit remains
close to the radio $L_*$ ($L_*$(radio)), 3CRR radio galaxies lie at
$\sim3-4L_*$(radio) 
(Willott et al. 2003) at all redshifts. Note also that 3CRR sources
are also the highest luminosity  
radio sources in the Hubble volume at $z<2$. 

\section{Results}\label{sec:results}

\begin{figure}
\centering
\ForceWidth{5in}
\hSlide{-1.8cm}
\vspace*{-1cm}
\BoxedEPSF{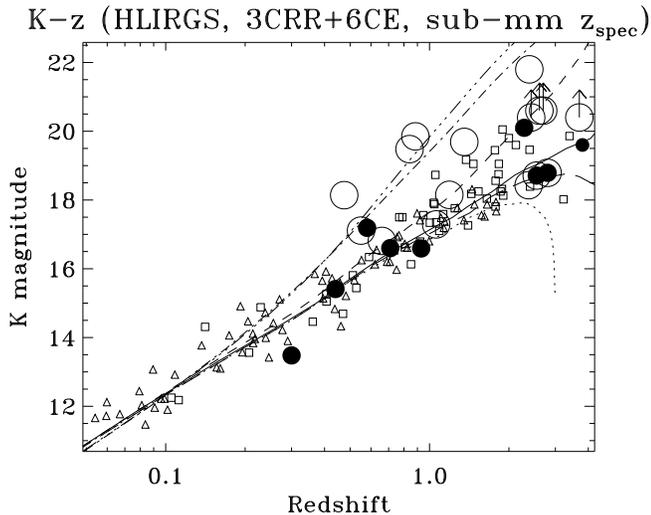}
\vspace*{-1cm}
\caption{\label{fig:kz1}$K$-band Hubble diagram of non-quasar
hyperluminous galaxies (filled circles), 3CRR radiogalaxies (open
triangles), 6CE radiogalaxies (open squares), and sub-mm selected
galaxies with spectroscopic redshifts (large open
circles). The 
hyperluminous radiogalaxy 4C41.17 (also the highest redshift HLIRG) is
plotted with a  
slightly smaller symbol, to distinguish it from the other HLIRGs. 
B2 0902+343 is excluded because it
is not clear whether the sub-mm flux is thermal, Downes et al. 1996.
Note that two sub-mm selected galaxies are also hyperluminous.
Also plotted are: a no evolution curve (short dash), instantaneous
starburst at $z=10$ followed by passive stellar evolution (full line),
$z=5$ starburst (long dash) and $z=3$ starburst, from Willott et
al. (2003). Also plotted are obscured starburst models of Takagi et
al. (2003): age $t/t_0=2$, compactness $\Theta=0.4$ (dash-dot) and
$\Theta=0.5$ (dash-dot-dot-dot). All models are for a
$3L_*(K)$ galaxy.}
\end{figure}

Figure \ref{fig:kz1} shows the $K-z$ relation of HLIRGs, compared to
that of bright radio sources from the 3CRR survey (Laing et al. 1983)
and 6CE survey 
(Eales et al. 1997). Willott et al. (2003) quotes a fit to the
radiogalaxy $K-z$ 
relation of 
\begin{equation}
K_{pred}=17.37 + 4.53\log_{10}(z) - 0.31(\log_{10}(z))^2 .
\end{equation}
3CRR has dispersion around this relation of only $\pm0.5$
magnitudes. The HLIRGs are statistically indistinguishable from the
3CRR radiogalaxies: a Kolmogorov-Smirnov test of the distributions of
$K-K_{pred}$ of HLIRGs and 3CRR radiogalaxies rejects the null
hypothesis of same distributions at only $23\%$ 
confidence.
\footnote{Although this test is only asymptotically
distribution-free, we 
verified the test in this case with a bootstrap analysis. We randomly 
selected $10$ 3CRR radiogalaxies, and compared $K-K_{pred}$ for this
subset against the remainder. By repeating this, we found that the
confidence levels returned  
are well-represented by a uniform distribution on the interval $[0,1]$
as required.} 
(We have excluded the hyperluminous radiogalaxy 4C41.17 from this test,
though its magnitude is consistent with the $K-z$ relation of other
HLIRGs table \ref{tab:kmags}). 

Curiously 
however, the same cannot be said of sub-mm selected galaxies, also
plotted in  
figure \ref{fig:kz1}. Even assuming the $K$-band limits are obtained
for these galaxies, the distributions are still dissimilar at $99.98\%$
confidence. Figure \ref{fig:kz2} shows the HLIRG $K-z$ relation in the
context of other potentially 
related populations. 
Note that the
photometric redshifts of three sub-mm selected galaxies and at least
$5$ with spectroscopic redshifts place them
securely away from the radiogalaxy locus. 

Neither the sub-mm galaxies nor the HLIRG samples from Rowan-Robinson
2000 are spectroscopically 
complete. Could this selection bias cause our HLIRG $K-z$ relation? In
a sample with a flux limit at two wavelengths, correlating the
luminosity at one wavelength with that at 
the other is essentially a distance vs. distance correlation
(e.g. Serjeant et al. 1998). However
this cannot be the cause of our HLIRG $K-z$ relation because 
(a) the apparent $K$ magnitudes 
are not well-represented by a $K$ flux limit: the magnitude histogram of
table \ref{tab:kmags} is more or less uniform from $K\sim13$ to
$K>20$; (b) more importantly, the optical selection ($R$ or $B$)
shows no greater evidence for clustering around a particular apparent
magnitude. If Malmquist bias (distance vs. distance correlations) 
were responsible, we would expect {\it more} clustering
around a particular apparent magnitude in the optical, compared to
the $K$-band. As discussed in Rowan-Robinson 2000,
there are probably 
optically-fainter HLIRGs still to be found in the IRAS database, and
our results make very specific predictions for their K magnitudes. 
However, if an as-yet-undiscovered selection effect results in the
undiscovered HLIRGs having fainter $K$ magnitudes, this would change
the results of this study. 

In both figure \ref{fig:kz1} and \ref{fig:kz2} we overplot passive
stellar evolution tracks of a $3L_*(K)$ galaxy
(where $L_*(K)$ is the K-band $L_*$) for an instantaneous starburst at
$z=10$, $z=5$, and $z=3$, as well as a no-evolution curve, as derived
by Willott et al. (2003), and two obscured starburst models of Takagi
et al. (2003).  We adopt $M_*(K)=-24.5$ in accordance with
the recent determination by Huang et al. (2003). We refer to the
passively evolving curve starting from a $z=10$ instantaneous
starbust, normalised to $M(K)=-24.5$ at zero redshift, as $K_*(z)$. 

\section{Discussion}\label{sec:discussion}

\begin{figure}
\centering
\ForceWidth{5in}
\hSlide{-1.8cm}
\vspace*{-1cm}
\BoxedEPSF{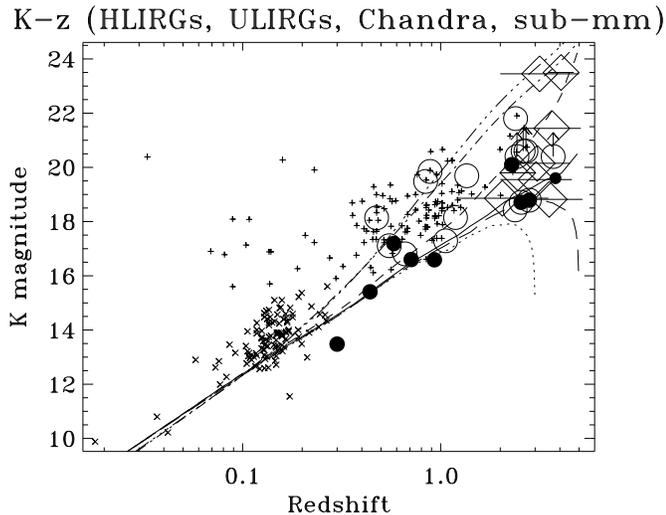}
\vspace*{-1cm}
\caption{\label{fig:kz2}
$K$-band Hubble diagram for HLIRGs (filled circles; 4C41.17 has
smaller symbol), sub-mm galaxies
with spectroscopic redshifts (large open circles) 
or photometric redshifts from detections in
$\geq3$ bands (large diamonds), 
non-quasar ultraluminous infrared galaxies ($\times$), 
and Chandra HDF-N $1$Ms non-quasar 
sources ($+$) from Barger et al. (2002). 
Chandra source photometry assumes $HK'-K=0.13+0.05(I-K)$ where
$I$-band photometry is available, and $HK'-K=0.3$ otherwise (Barger et
al. 2002). 
For clarity we omit the 
radiogalaxies. The same models are plotted as figure \ref{fig:kz1}. 
}
\end{figure}

The difference between the HLIRG and sub-mm galaxy $K-z$ relations is
all the more puzzling given our discovery that some HLIRGs would have
comparable sub-mm fluxes to sub-mm-selected galaxies, if redshifted to
$z\sim3$ (Farrah et al. 2002b), including two of the HLIRGs in the
present paper. Given a template spectral energy
distribution with a sufficiently warm colour temperature, many of the
sub-mm-selected galaxies could have bolometric luminosities
approaching $10^{13}L_\odot$. 
Five of the ten HLIRGs in this paper were selected at
$60\mu$m, and two of the remainder were discovered by 
follow-ups of radiogalaxies, either of which may have introduced a
bias towards high AGN bolometric 
contributions. 
If so, this would suggest that the position on the
$K-z$ relation may be related to the presence of the most massive
AGN.  A prediction of this interpretation is that the multi-wavelength
data on sub-mm galaxies from SIRTF (e.g. Lonsdale et al. 2003) should
find an anticorrelation between $K-K_*$ magnitudes and the bolometric
contributions in the mid-infrared.

Alternatively, the HLIRG $K-z$ relation may be intrinsic, rather than
the product of a subtle AGN bias. 
All of the galaxies currently {\it proved} to be hyperluminous
currently lie on a tight $K-z$ relation, including those selected in
the sub-mm. 
Also, radiative transfer models of the Rowan-Robinson (2000) HLIRGs do
not find the AGN to be bolometrically dominant in all cases
(e.g. Verma et al. 2002, Farrah et al. 2002b). 
There is
no prima facie reason to suppose that HLIRGs should necessarily follow
the radiogalaxy $K-z$ relation. 
For example, the 
obscured high-redshift AGN detected by Chandra are not found with similar
$K$ magnitudes to radiogalaxies (or HLIRGs, figure \ref{fig:kz2}),
which Willott et al. (2001) 
argued was due to the Chandra sources hosting smaller mass nuclear black
holes. 
However, the relative number densities
of radiogalaxies and HLIRGs lend plausibility to a physical link
between the populations, or at least a common progenitor. 
Rowan-Robinson (2000) lists $16$ HLIRGs with $z\leq1$, implying a
lower limit to the $z\leq1$ HLIRG number density of
$\geq3\times10^{-10}h^3$Mpc$^{-3}$, and also 
estimates
that only $\sim10-20\%$ of HLIRGs have been identified to date.
These number densities are comparable to the 
$z\leq1$ number
density of 3CRR radio galaxies ($1.1\times10^{-9}h^3$Mpc$^{-3}$) and
significant compared to $z\leq1$ 3CRR active galaxies as a whole
($2.5\times10^{-9}h^3$Mpc$^{-3}$). Notably, both HLIRGs and 3CRR are
the most luminous in the Hubble volume of their class. 

This does not necessarily imply that radiogalaxies and HLIRGs must be
the most massive galactic systems in the Hubble volume. Huang et
al. (2003) measured the space density of $10L_*(K)$ galaxies 
to be
$2.1\times10^{-8}h^3$Mpc$^{-3}$ at $z<0.4$. 
These galaxies are $2.5$ magnitudes brighter than the HLIRG and
radiogalaxy host galaxies, and $\sim10\times$ more numerous than
radio-loud AGN. However, the space density of $>10L_*(K)$ galaxies has
not been determined at higher redshifts. 

\begin{figure}
\centering
\ForceWidth{5in}
\hSlide{-1.8cm}
\vspace*{-2cm}
\BoxedEPSF{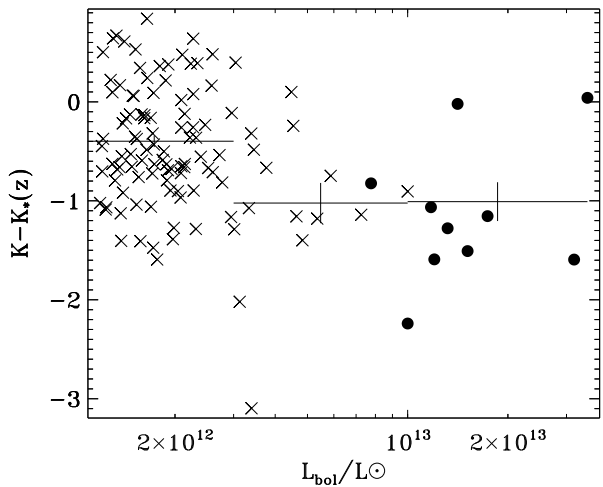}
\vspace*{-1cm}
\caption{\label{fig:kz3}
$K$-band luminosity relative to $L*$ ($z=10$ starburst model), plotted
against bolometric 
luminosity. HLIRGs: filled circles; ultraluminous galaxies: crosses. 
Also plotted are the mean $K-K_*$ values in binned luminosity ranges
(horizontal lines). 
The standard deviation is $\sim\pm0.6$ magnitudes in each bin, and the
errors on the means are plotted as vertical lines. 
Note that for our cosmology, some of the Rowan-Robinson (2000)
compilation slip just below the HLIRG threshold, and one of
the ultraluminous galaxies becomes HLIRGs. For consistency with
previous work, we restrict our HLIRGs to the Rowan-Robinson (2000)
compilation, but as can be seen from this figure this choice does not
affect our conclusions. 
}
\end{figure}

There is 
evidence in figure \ref{fig:kz3} that infrared luminosity scales with
host luminosity, in a manner reminiscent of the trend of host
luminosity with radio lobe luminosity in radiogalaxies (Willott et
al. 2003). The difference in the mean $K-K_*$ for the
$1-3\times10^{12}L_\odot$ and $>10^{13}L_\odot$ bins is significant
at $99.6\%$ confidence using Student's T statistic, though the objects
in these bins span 
different redshifts, so differential evolution may also be a factor
(e.g. Serjeant et al. 1998). Samples of infrared-luminous galaxies
spanning a narrow range in redshift, but a wide range in luminosity,
are needed to test this relationship definitively. Such samples may
become available with the advent of SIRTF. 

If radiogalaxy and HLIRG activity is both short lived and rare (in the
sense that a $5L_*(K)$ galaxy has a low probability of ever hosting a
radiogalaxy or HLIRG), then the lack of $>10L_*(K)$ HLIRGs might be
explained by the finite size of the volume surveyed so far for 
HLIRGs. 
If we assume that $K$-band galaxies in the range $0.5-5L_*(K)$ are the
hosts of short-lived HLIRG activity with $10^{13-14}L_\odot$, and we
interpolate from the  
$K$-band luminosity function (Huang et al. 2003) 
keeping the HLIRG duty cycle constant, 
then the space density of $10^{14-15}L_\odot$ galaxies should be
around a factor of $\sim1000$ lower. 
This is, of course, only a toy model: the $K$-band
luminosity function is unlikely to 
keep the same shape at all redshifts, the luminosities may not scale
with the host galaxy masses, and the duty cycle assumptions
may be ill-founded, but this does raise the interesting question of
the existence of still more extreme 
populations of infrared galaxies. 
Source count models differ widely in their predictions at these
luminosities (Pearson 2001, Rowan-Robinson 2001). 
Whether such systems do in fact exist may be
testable with the next generation of sub-mm/mm-wave survey
facilities, such as SCUBA-2 (Holland et al. 2003) or the LMT (Baars et
al. 1998). 


In short, there are plausible precedents for
abundant populations of galaxies with evidence of dust-enshrouded AGN,
extreme luminosities and/or large stellar masses, which are 
many times more luminous in $K$, or 
many times less luminous, than HLIRGs. 
The fact that these populations
do not host HLIRG activity suggests 
that the similarity of the HLIRG and radiogalaxy 
$K-z$ relations is due to a direct physical link between the two
phenomena, such as an evolutionary connection, or a common progenitor
population. 
Alternatively, the position of HLIRGs on the $K-z$ relation may be
solely related to the presence of the most massive AGN (see above),
for which a key test is whether the SIRTF detects hyperluminous
AGN activity in sub-mm selected galaxies.

Surprisingly, sub-mm galaxies have a very different distribution in
the $K-z$ plane. 
This is contrary to the result of Dunlop (2002) 
due to the subsequent increase in (ostensibly) reliable spectroscopic
redshifts.  
The spectroscopic redshift of ELAIS N2850.1 
is anomalous compared to its radio:sub-mm ratio, which led Chapman et
al. (2002) to suggest that the system may be lensed. Placing the
system at higher redshift may well restore ELAIS N2850.1 to closer to
the locus of the radiogalaxy $K-z$ relation. 
Nevertheless, several photometric or spectroscopic 
redshifts place sub-mm
galaxies away from 
the radiogalaxy $K-z$ relation (figure \ref{fig:kz2}). 
Plausibly, these may represent separate populations; there is no
reason to suppose that sub-mm selected galaxies represent a single
homogeneous population of objects (e.g. Dannerbauer et al. 2002), as
with Extremely Red Objects (e.g. Smail et al. 2002a). 

One possibility is that some sub-mm galaxies are less massive systems;
another is that 
not all of them  
are the progenitors of the most massive ellipticals
(Efstathiou \& Rowan-Robinson 2003,
Kaviani et al. 2003) but rather are cool cirrus-dominated objects. 
Alternatively, some sub-mm galaxies may be heavily extinguished even
in the observed-frame $K$-band. 
Such an interpretation is physically plausible: figures \ref{fig:kz1}
and \ref{fig:kz2}  
show the predictions of such a model from 
Takagi et al. 2003.

The HLIRG $K-z$ relation could be used to estimate redshifts of
HLIRG candidates, as was the early 
practice for radiogalaxies (e.g. Dunlop \&
Peacock 1990). Based on the 3CRR radiogalaxy $K-z$ relation, the $K$
magnitude is sufficient to determine the redshift of HLIRGs 
to better than $\pm10\%$ in $(1+z)$ at all redshifts, provided the
systems are indeed hyperluminous and are not quasars. Regarding 
the hyperluminous quasars in Farrah et al.
(2003), we can predict that the
hyperluminous quasar LBQS 1220+0939 should be 
dominated by the host galaxy
flux in $K$, while the hyperluminous quasar 
IRAS F10119+1429 should be 
dominated by the nuclear component in $K$. The remaining cases in
Farrah et al. (2003) should be intermediate between these two cases. 

Only two of the known non-quasar HLIRGs lack $K$-band photometry, so
further tests of the HLIRG $K-z$ relation using HLIRGs discovered to
date must rely on 
sub-arcsecond near-infrared imaging of hyperluminous quasar hosts. 
Alternatively, both SIRTF and ASTRO-F will have sensitive $L$ and $M$-band
cameras, which can further reduce the contribution from the active
nucleus by sampling closer to the rest-frame $K$-band. 

\section{Acknowledgements}
We would like to thank Chris Willott for kindly supplying the 
stellar evolution curves, and the anonymous referee for very 
helpful
suggestions, including the possibility that an AGN bias in the
Rowan-Robinson (2000) sample may be related to their position on the
$K-z$ relation. 
This research was supported by PPARC grant PPA/G/0/2001/00116, and
Nuffield Foundation grant NAL/00529/G. 
This publication makes use of data products from the Two Micron All
Sky Survey, which is a joint project of the University of
Massachusetts and the Infrared Processing and Analysis
Center/California Institute of Technology, funded by the National
Aeronautics and Space Administration and the National Science
Foundation. This research made use of the NASA/IPAC Extragalactic
Database (NED) which is operated by the Jet Propulsion Laboratory,
California Institute of Technology, under contract with the National
Aeronautics and Space Administration.

\end{document}